# ROSAT RESULTS ON GROUPS AND CLUSTERS OF GALAXIES[1]

## Sabine Schindler

Max-Planck-Institut für extraterrestrische Physik, 85748 Garching, Germany

Max-Planck-Institut für Astrophysik, 85748 Garching, Germany

**Summary.** – We present a selection of ROSAT results on the morphology, the evolution, and the masses of galaxy clusters and groups. ROSAT provides the first complete X-ray view on the Virgo cluster: it reveals a complex structure similar to the optical appearance. Indications for ongoing evolution is found in a number of clusters traced both by the surface brightness distribution and temperature variations. With ROSAT data, which provide spatial and spectral resolution at the same time, good mass estimates can be obtained for various systems from small groups to very massive clusters covering a mass range of a few $10^{13}$ to $4.6 \cdot 10^{15} \mathcal{M}_\odot$. The gas to dark matter ratio for clusters is 10-30% while for groups slightly smaller values (3-25%) are found. The most luminous ROSAT cluster RXJ1347.5-1145 is presented which is an ideal object to compare the X-ray mass with the mass derived by gravitational lensing.

## 1 Introduction

ROSAT is an instrument well suited for the study of clusters of galaxies because of its high sensitivity and its high spatial resolution with simultaneous energy resolution.

With ROSAT observations a number of astrophysical and cosmological questions can be addressed ranging from the dark matter distribution within clusters to the large-scale mass distribution in the universe. From the X-ray luminosity function and the evolution of clusters cosmological parameters can be inferred.

The talk focuses on three aspects of ROSAT research on clusters and groups: X-ray morphology, evolution and mass determination.

## 2 X-ray Morphology

ROSAT provides by far the most detailed view on the X-ray morphology of clusters, e.g. it gives the first complete X-ray view of the closest cluster – the Virgo cluster (Böhringer et al. 1994). Figure 1 shows that the hot luminous gas is extending over most of the optically visible cluster. The X-ray structure of the Virgo cluster is complex with several subclusters around M87, M86, M49 and M60, indicating that the cluster is still evolving.

---

[1] To appear in "Interacting Galaxies: in Pairs, Groups and Clusters", Sant' Agata sui due Golfi, September 12-15, 1995, *Astrophysical Letters and Communications*

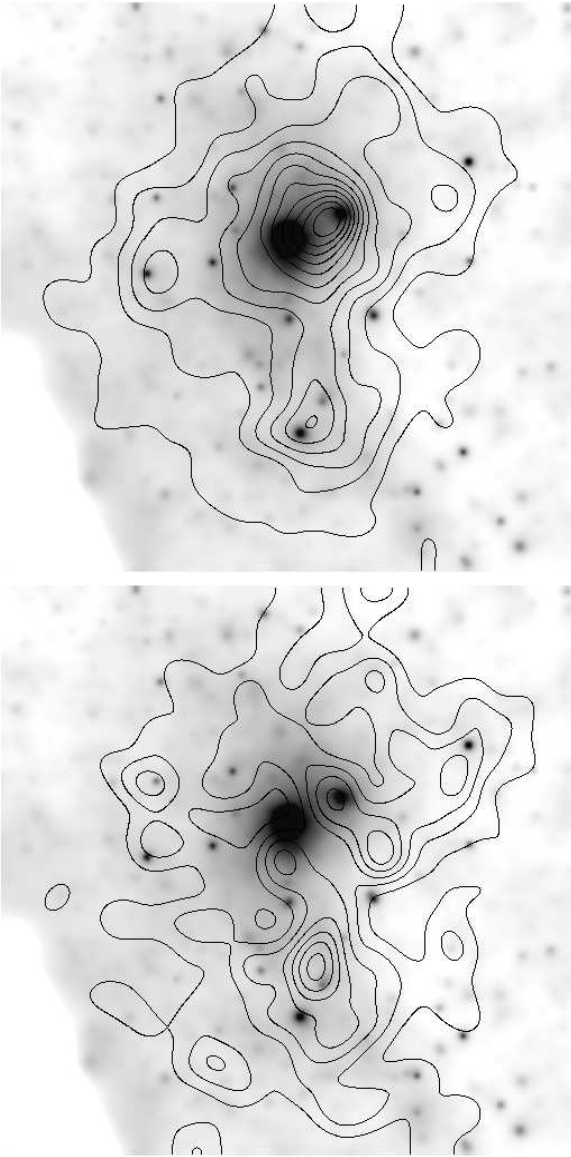

Figure 1: X-ray emission of the intra-cluster gas (colour image) from the Virgo cluster in the ROSAT All Sky Survey in the hard band (0.4-2.4 keV) (Böhringer et al. 1994). Superposed on the X-ray image are contours of galaxy number density (a) of the 1300 member galaxies and (b) of only the late-type galaxies of the Virgo Cluster Catalog (Binggeli et al. 1985). The images have a size of 12.8°×12.8°. The X-ray image is smoothed with a variable Gaussian filter with $\sigma = 24$ arcmin on the faintest levels and decreasing filter size with increasing surface brightness. The galaxy distribution is smoothed with a 2-dimensional Gaussian with $\sigma = 24$ arcmin.

Superposed on the X-ray image are contours of constant galaxy number density taken from the Virgo Cluster Catalog (Binggeli et al. 1985). There is good correspondence between the two components: the galaxy distribution shows a similar structure with several subclumps at about the same positions. This similarity implies that both components trace the same potential and are almost in equilibrium in it.

In Fig.1b the number density of only the late-type galaxies (spirals, irregulars, Im, BCD) is plotted. They show a completely different distribution. The galaxy density maxima do not correspond with the X-ray maxima. The late-type galaxies are much less concentrated than the early types, in accord with the density-morphology relation (Dressler 1980) and the radius-morphology relation (Whitmore et al. 1993).

Not only rich clusters but also some groups of galaxies were found to have X-ray emission. In 8 out of the 97 Hickson Compact Groups X-ray emission was detected in the ROSAT All Sky Survey (Ebeling et al. 1994). Some of them show clearly extended emission. Out of the 40 loose groups from the CfA Redshift Survey (Ramella et al. 1989) at least 2 groups (RGH12 and RGH80) have X-ray emission (Böhringer & Ramella, in preparation). Both, the compact and the loose groups have X-ray temperatures around 1 keV.

## 3  Cluster Evolution

In numerical simulations of cluster evolution it was shown that the dynamical state of a cluster can best be deduced from the temperature distribution of the intra-cluster medium (Schindler & Müller 1993). ROSAT with its simultaneous spatial and spectral resolution allows for a determination of the gas temperature in different regions for a number of clusters.

A cluster in which the merger hypothesis deduced from the ROSAT morphology (Briel et al. 1991) could be confirmed by an X-ray temperature map is A2256 (Briel & Henry 1994). The map shows regions of different temperature implying that the cluster is not in equilibrium.

Another example is the bimodal cluster A3528 (Fig.2, Schindler 1995a) in the Shapley Supercluster. There are not enough counts in the cluster to deduce a temperature map, but each subcluster could be divided into two semicircles – one facing the other subcluster, the other looking in the opposite direction. We found a trend that the temperature in the two inner semicircles is higher than in the outer ones. This temperature trend is an indication that the two subclusters are not a chance alignment but will collide within the next few $10^8$ years.

Unfortunately, such a temperature analysis is not possible for many clusters. Therefore a new method for the quantification of substructure from X-ray images alone was developed. The amount of substructure in clusters is a measure for the age of the clusters, from which cosmological parameters like the mean density in the universe $\Omega_0$ can be inferred. Based on a method for optical images (Bender & Möllenhoff 1987) Neumann & Böhringer (1995a) developed a technique that fits ellipses to different isophote levels i.e. radii of X-ray clusters. The fits yield parameters like the major axis, eccentricity, and position angle. The variation of these parameters with radius is a measure for the substructure in the cluster.

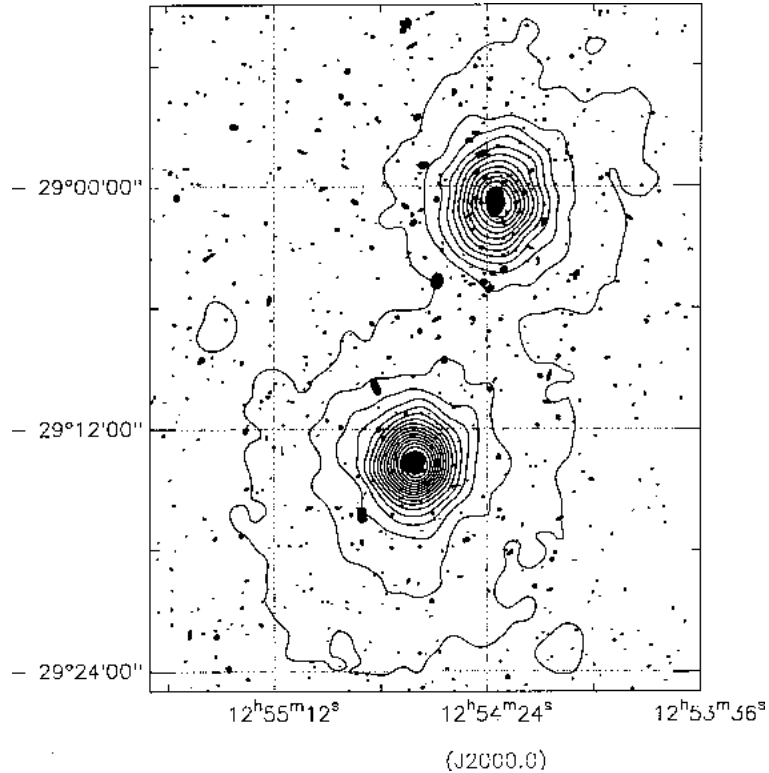

Figure 2: ROSAT PSPC image of the cluster A3528 in the hard band (0.5-2.0 keV). The double structure of the X-ray image is clearly visible.

Recently, a cluster was found where the merging process can be studied closeby. The nearby cluster A3627 (z=0.016) was found by Kraan-Korteweg et al. (1995) to be an optically very rich cluster. Although it was overlooked in all X-ray surveys it turned out to be the $6^{th}$ brightest X-ray cluster (Böhringer et al. 1995). Fig. 3a shows a ROSAT/PSPC image of A3627. The contours are not spherically symmetric but show an elongation in northwest – southeast direction. When subtracting a spherically symmetric image from Fig. 3a – constructed with the parameters of the main component in the west – a compact subcluster shows up in the southeast (Fig. 3b). Obviously, the merging process has progressed already to a stage where the infalling subcluster is being distorted. For A3627 a mass around $10^{15}\mathcal{M}_\odot$ was found, that is at least half the mass of the prominent Perseus and Coma clusters. A first temperature analysis shows temperature differences between the main component and the subcluster, therefore this nearby and X-ray bright cluster should be one of the most interesting targets to study cluster mergers with spatially resolved X-ray spectroscopy.

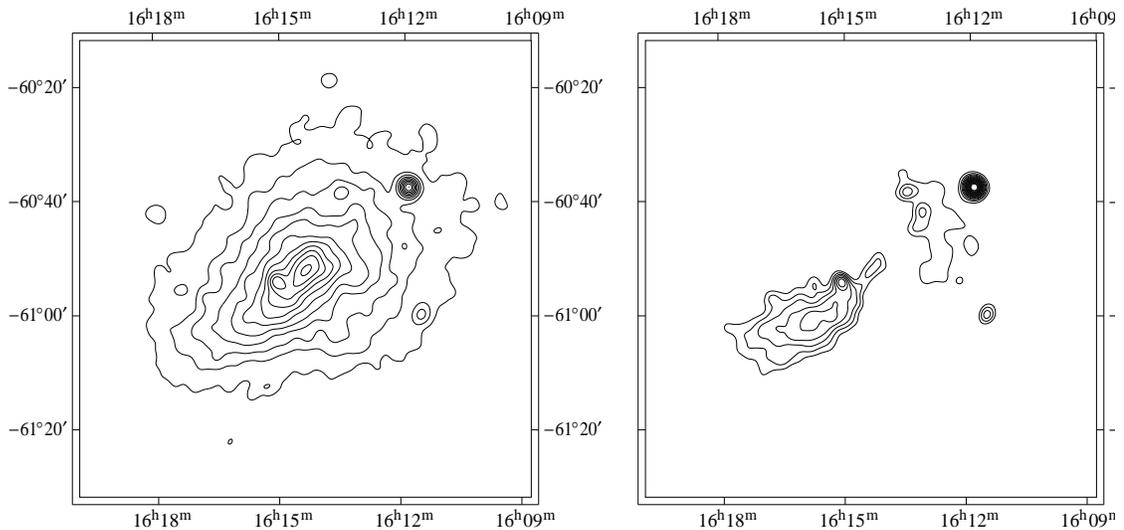

Figure 3: a) ROSAT/PSPC image of the $6^{th}$ brightest X-ray cluster A3627 in the hard ROSAT band (0.5-2.0 keV). The image is smoothed with a Gaussian with $\sigma = 1$ arcmin. The contour spacing is $9.6 \cdot 10^{-4}$cts/s/arcmin$^2$ with the lowest contour being at $1.6 \cdot 10^{-3}$cts/s/arcmin$^2$. b) After subtracting the main component a compact subcluster is visible which is falling into the main cluster from the west.

## 4 Mass Determination

With an accurate determination of the masses of clusters which are the most massive, clearly defined objects in the universe cosmological parameters like $\Omega_0$ or the baryon to total mass fraction can be constraint.

With the assumption of hydrostatic equilibrium and spherical symmetry the equation for the integrated mass can be deduced

$$M(r) \propto rT \left( \frac{d \ln \rho}{d \ln r} + \frac{d \ln T}{d \ln r} \right) \quad (1)$$

where $\rho$ and $T$ are the density and the temperature of the intra-cluster gas. ROSAT observations provide both required quantities – the density and the temperature distribution. A summary of total masses and gas mass fractions derived from a sample of deeply observed and well analysed clusters and groups of galaxies is given in Table 1. As the listed gas mass fractions of 5-30% provide a lower limit for the baryon to total mass ratio in the universe (assuming that the potentials of the clusters accumulate mass indiscriminately) there seems to be a discrepancy with primordial nucleosynthesis.

A problem for mass determination from ROSAT data is that the temperature is often not very well constraint and has errors up to 50%. For the transformation of these temperature errors to errors in the mass profile a Monte-Carlo method was developed by Neumann & Böhringer (1995b). They calculate about 10000 possible temperature profiles within the allowed temperature range. With these temperature profiles the range of the mass profile is defined.

Table 1. Summary of masses and composition of clusters and groups of galaxies (from Böhringer 1995).

|  | rich clusters | groups of galaxies |
|---|---|---|
| $M_{grav}$ | $\approx 0.5 - 5 \cdot 10^{15} \mathcal{M}_\odot$ | $\approx 0.3 - 5 \cdot 10^{14} \mathcal{M}_\odot$ |
| galaxies | 2-7% $h_{50}^{-1}$ of $M_{grav}$ | 5-10% $h_{50}^{-1}$ of $M_{grav}$ |
| intracluster gas | 10-30% $h_{50}^{-1}$ of $M_{grav}$ | 3-25% $h_{50}^{-1}$ of $M_{grav}$ |
| dark matter | 60-85% $h_{50}^{-1}$ of $M_{grav}$ | 60-90% $h_{50}^{-1}$ of $M_{grav}$ |
| iron abundance | $0.35 \pm 0.15$ solar | $0.25 \pm 0.15$ solar |

Recently, in few clusters a discrepancy between the X-ray mass and the mass determined by the gravitational lensing effect has shown up. Therefore the question has been raised whether the above mentioned two assumptions – hydrostatic equilibrium and spherical symmetry – are good assumptions. To check the reliability of mass determination by X-ray data we make use of combined N-body/hydrodynamic models (Schindler & Müller 1993). We compare the true mass of the model with the mass determined from a simulated observation of the model cluster (Schindler 1995b). For some extreme cases, which show either strong shocks (deviation from hydrostatic equilibrium) or pronounced substructure (deviation from spherical symmetry) we find easily mass deviations of 100%, where shocks typically cause an overestimation while substructure leads to an underestimation of the mass. For all the roughly virialised clusters, on which the mass determination is usually applied (even some of the substructure cases can be reduced to these kind of clusters by clipping the critical regions from the cluster) we find a typical deviation from the true mass of 15% without any systematic over- or underestimation. As this error is smaller than today's observational errors the mass determination by X-ray data can be regarded as very reliable.

An ideal example where both mass determination methods can be applied was found recently in an ESO Key Programme which performs a redshift survey of ROSAT detected clusters (Böhringer 1994; Guzzo et al. 1995). The cluster RXJ1347.5-1145 at z=0.451 was found to be the most luminous cluster yet discovered in the ROSAT band (0.1 - 2.4 keV) with a luminosity of $(6.2 \pm 0.6) \cdot 10^{45}$ erg s$^{-1}$ (Schindler et al. 1995). We are going to derive a very accurate mass estimate with our follow-up observations with the ROSAT/HRI and ASCA. Furthermore, two bright arcs were discovered in this cluster (Fig.4). They are opposite to each other with respect to the cluster centre and have a distance from it of about 35″ (= $240 h_{50}^{-1}$ kpc), a configuration and a radius that enables the probing of the mass in a rather large cluster volume. As soon as the redshifts of the arcs are measured we can compare the mass estimates by the two different methods. Due to the extremely high luminosity of RXJ1347.5-1145 an exceptional high mass is expected which would be a challenge for standard cosmological models.

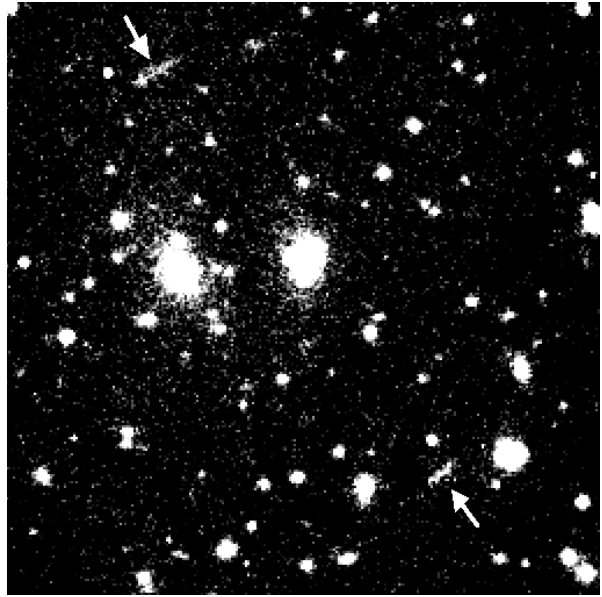

Figure 4: R image of the centre of the cluster RXJ1347.5-1145 over a square field with 1.4 arcmin side slightly smoothed with a Gaussian with $\sigma=0.5$pixel. The two arcs are northeast and southwest of the central galaxy (North is up, East is left).

## Acknowledgements

I thank the ROSAT team for preparing the data and providing the analysis software. I am grateful to my colleagues Hans Böhringer and Doris Neumann for many fruitful discussions. The generous hospitality of the Astronomical Institute of the University of Basel is greatly appreciated. I acknowledge financial support by the Verbundforschung.

**Question from Sancisi:** In RXJ1345-1145, how do the mass estimates from X-rays and from gravitational lenses compare?

**Answer:** A very preliminary comparison with estimated redshifts of the arcs and a simple lens model assuming a spherically symmetric lens yields a lensing mass larger than the X-ray mass.

**Question from Burbidge:** Many of the extended X-ray sources in clusters are centered on a central cD galaxy or a giant elliptical. This is clearly the case for M87 and other galaxies you showed. Now these galaxies are often very powerful radio sources and are active in other ways. Thus I wonder why it cannot be interpreted in terms of all of the energy input for the hot gas coming from the active galaxy, not always from gravity through a "cooling flow"?

**Answer:** The required energy to heat the intra-cluster gas is about $10^{64}$ ergs which is extreme for a single radio source. Furthermore, it is not easy to explain the extended and often structured x-ray emission by one source located in the centre of the cluster.

**Question from Ohashi:** Is there a Seyfert galaxy lying in the centre of A3627? If not, is there a cD galaxy in the centre of this cluster?

**Answer:** The central galaxy is a WAT radio galaxy (PKS1610-60). It is located between the main X-ray component and the subcluster.

**Comment from Ohashi:** The estimated mass of the hot gas in poor groups is often comparable or larger than the total mass of galaxies. Therefore, it seems impossible to explain the X-ray emission without invoking the diffuse hot gas.

**Two Comments from Ramella:** The first on Burbidge's question: it is true that we see X-rays mostly from the brightest ellipticals in groups. We only have two observations long enough to see that there is an extended emission around the galaxies. I think in the loose groups we do not see in the X-rays the underlying large scale gravitational potential of the group. In fact dell'Antonio et al. 1994 show that $L_X$ versus $\sigma$ is flatter for groups than for clusters, even if one removes the elliptical galaxy. The $L_X$ versus $\sigma$ slope is consistent with $L_X \propto N_{galaxies}$ being the dominant part of the correlation.

The second comment is that if one superimposes the X-ray maps of HCG42 and HCG97 of the X-ray maps of RGH12 and RGH80 they are almost undistinguishable.